\documentclass{article}
\usepackage{spconf,amsmath,graphicx}

\usepackage{multirow}  
\usepackage{booktabs}
\usepackage{booktabs,multirow}
\usepackage{tabularx}
\usepackage{ragged2e} 
\usepackage{float}

\title{LSU-Net: Lightweight Automatic Organs Segmentation Network For Medical Images}
\name{Yujie Ding\textsuperscript{1}, Shenghua Teng\textsuperscript{1}, Zuoyong Li\textsuperscript{2}, Xiao Chen\textsuperscript{2}$^*$
\thanks{*Xiao Chen is the corresponding author.}
\thanks{This work was supported in part by the National Natural Science Foundation of China under Grant 62471207, in part by the Natural Science Foundation of Fujian Province under Grant 2024J02029, in part by the Joint Funds for The Innovation of Science and Technology in Fujian province under Grant 2023Y9280, in part by the Open Project of Fujian Key Laboratory of Medical Big Data Engineering under Grant KLKF202301, in part by the Special Fund of Minjiang University under Grant MJY22019.}}

\vspace{-1em} 
\address{\textsuperscript{1} College of Electronic and Information Engineering,\\
Shandong University of Science and Technology, Qingdao 266590, China\\
\textsuperscript{2} Fujian Provincial Key Laboratory of Information Processing and Intelligent Control, \\
School of Computer and Big Data, Minjiang University, Fuzhou 350121, China}
\vspace{-1em} 

\usepackage{hyperref}
\hypersetup{
 	colorlinks=true,
 	pdfstartview=Fit,
	breaklinks=true,
        linkcolor=blue,
        anchorcolor=blue,
        citecolor=blue,
        urlcolor=blue}

\begin{document}

%
\maketitle

\begin{abstract}
UNet and its variants have widespread applications in medical image segmentation. However, the substantial number of parameters and computational complexity of these models make them less suitable for use in clinical settings with limited computational resources. To address this limitation, we propose a novel Lightweight Shift U-Net (LSU-Net). We integrate the Light Conv Block and the Tokenized Shift Block in a lightweight manner, combining them with a dynamic weight multi-loss design for efficient dynamic weight allocation. The Light Conv Block effectively captures features with a low parameter count by combining standard convolutions with depthwise separable convolutions. The Tokenized Shift Block optimizes feature representation by shifting and capturing deep features through a combination of the Spatial Shift Block and depthwise separable convolutions. Dynamic adjustment of the loss weights at each layer approaches the optimal solution and enhances training stability. We validated LSU-Net on the UWMGI and MSD Colon datasets, and experimental results demonstrate that LSU-Net outperforms most state-of-the-art segmentation architectures.
\end{abstract}
\section{INTRODUCTION}
\label{sec:intro}
Medical image segmentation is crucial for extracting structural details to aid diagnosis and treatment. Deep learning has significantly advanced the field with diverse network architectures, especially the U-shaped encoder-decoder design starting with UNet \cite{UNet}. Subsequent models, such as UNet++ \cite{UNet++}, Attention U-Net \cite{Attention-UNet}, UNet3+ \cite{UNet3+}, Rolling-UNet \cite{RollingUNet}, and UNeXt \cite{UNext}, introduced innovations like nested structures and dense skip connections, further improving performance.

Inspired by the success of Transformers in NLP and computer vision, Transformer-based networks like Swin-UNet \cite{SwinUNet} and TransUNet \cite{TransUNet} capture long-range dependencies for segmentation. Models such as MedT \cite{MedTrans} and TransBTS \cite{TransBTS} enhance accuracy with advanced attention mechanisms \cite{Atten2, Atten3}.

Despite these advances, many models overlook the computational constraints of real-world medical settings. Previous research also struggles to balance parameters and performance within these limits. To address these challenges, we propose LSU-Net, a lightweight segmentation model performing multi-scale supervised learning while maintaining strong segmentation performance. LSU-Net retains the classic encoder-decoder architecture, applying different modules at various stages. In shallow stages, it employs low-parameter convolutional modules to capture features efficiently. In deep stages, LSU-Net uses MLP to shift deep features, capturing regional dependencies and optimizing feature representation. We also implement horizontal deep supervision, dynamically adjusting loss weights to compare multi-scale information.

To validate the effectiveness of LSU-Net, we evaluated it on the UWMGI \cite{uw-madison-gi-tract-image-segmentation} and MSD Colon \cite{MSD} datasets. The results show that our method achieves excellent segmentation performance with fewer parameters, demonstrating its applicability in real-world environments.

\section{METHOD}
\label{sec:method}
This section provides detailed information about LSU-Net, including its overall network architecture, core components, and loss functions. The core components discussed include the Light Conv Block and the Tokenized Shift Block.

\begin{figure*}[h!]
    \centering
    \includegraphics[width=0.97\linewidth, trim=0.5cm 0.5cm 0.5cm 0.8cm, clip]{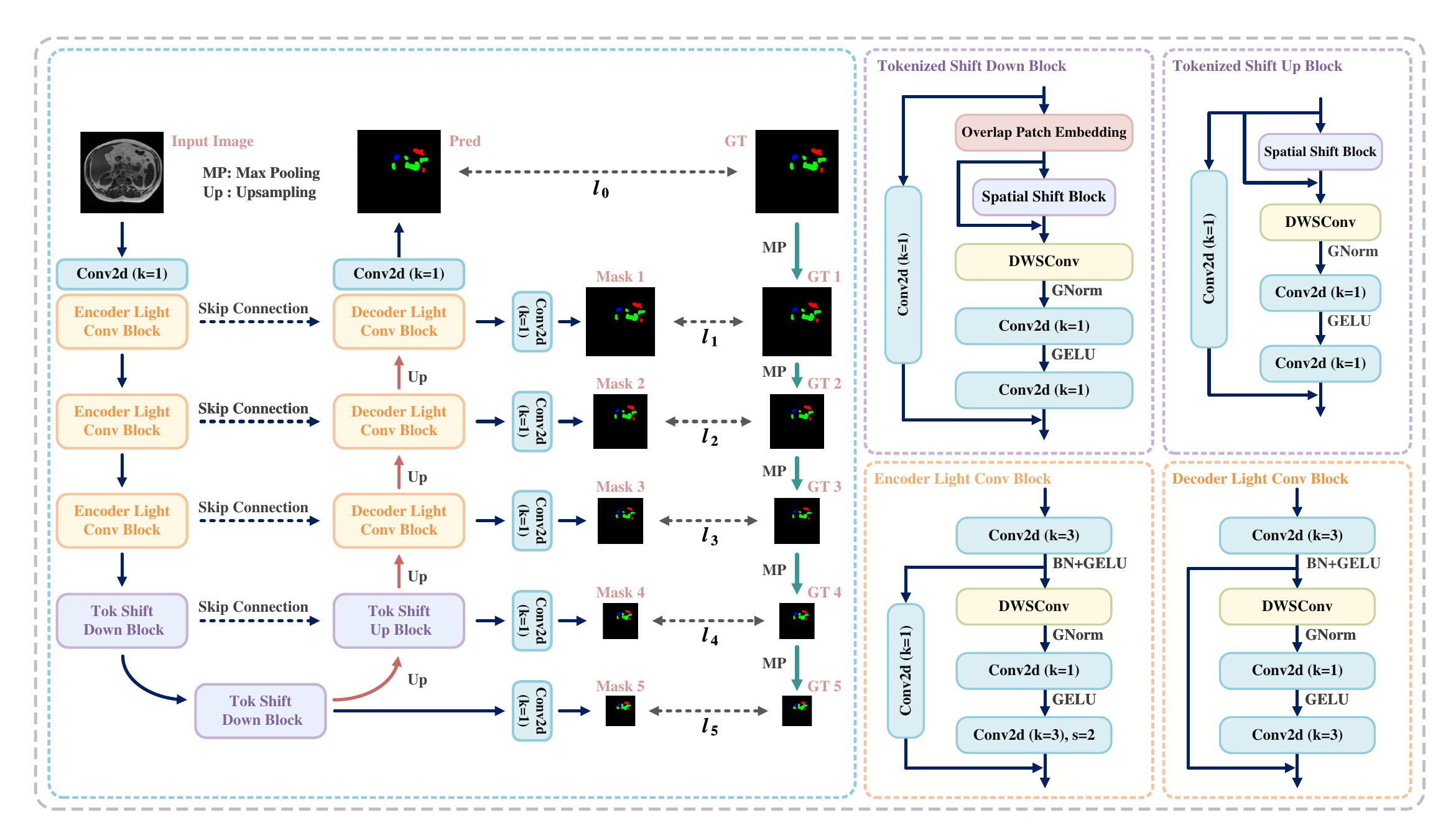}
    \caption{The overview of the Lightweight Shift U-Net(LSU-Net). The downsampling in the encoder is achieved by setting the convolution stride to 2. Deep Supervision is used at each layer. \(l_{0}\) to \(l_{5}\) means loss function in each level. The upper right of the figure shows the Tokenized Shift Block, while the lower right of the figure shows the Light Conv Block.}
    \label{fig:LSU-Net}
\end{figure*}

\noindent \textbf{The Overall Architecture.} We propose LSU-Net, a lightwei-
\noindent ght network designed for abdominal organ segmentation through efficient feature extraction and processing. LSU-Net follows a U-shaped architecture, as shown in \textbf{Figure} \ref{fig:LSU-Net}. To a-

\noindent chieve this, we introduce two modules: the Light Conv Block and the Tokenized Shift Block. The encoder includes the Encoder Light Conv Block and Tokenized Shift Down Block, while the decoder comprises the Decoder Light Conv Block and Tokenized Shift Up Block.


\noindent \textbf{Light Conv Block.} The Light Conv Block, as shown in the bottom left of \textbf{Figure} \ref{fig:LSU-Net}, consists of several layers: First, the main branch undergoes a 3×3 convolution to extract local features and capture spatial information using a 3×3 kernel. Next, Batch Normalization (BN) and GELU activation functions enhance feature representation, as shown in Eq.\eqref{eq:1}:
\begin{equation}
X_1 = GELU\left(BN\left({Conv}_{3 \times 3}\left(X\right)\right)\right)
\label{eq:1}
\end{equation}
where \(X\) represents the input and \(X_{1}\) denotes the output of this stage.

Following this, Depthwise Separable Convolution (DWSConv) is used for feature extraction, significantly reducing computation and parameters. Next, Group Normalization (GNorm) is applied. A 1×1 convolution then adjusts the number of channels in the feature map, and GELU activation enhances non-linear representation. Finally, a 1×1 convolution with a stride of 2 downsamples the spatial dimensions of the feature map in the encoder, while a 1×1 convolution with a stride of 1 is applied in the decoder:
\setlength{\belowdisplayshortskip}{8pt}  
\begin{equation}    
X_{2} = {Conv}_{1 \times 1}\left( GNorm\left( DWSConv\left( X_{1} \right) \right) \right)
    \label{eq:2}
\end{equation}
\begin{equation} 
X_{3} = {Conv}_{1 \times 1}\left( {GELU\left( X \right.}_{2} \right))
    \label{eq:3}
\end{equation}

Additionally, the Conv Block includes various skip connections. The encoder employs a 1×1 convolution as a skip connection to alter channel dimensions. The outputs of the encoder and decoder are given in Eq.\eqref{eq:4} and Eq.\eqref{eq:5}, respectively. This design integrates various convolutional and normalization operations to balance feature extraction, computational efficiency, and model performance.
\begin{equation} 
Y_{encoder} = X_{3} + {Conv}_{1 \times 1}(X_{1})
    \label{eq:4}
\end{equation}
\begin{equation} 
Y_{decoder} = X_{3} + X_{1}
    \label{eq:5}
\end{equation}

\begin{figure}[H]
    \centering
    \resizebox{0.92\columnwidth}{!}{%
        \includegraphics[trim=0cm 0.6cm 0cm 0.5cm, clip]{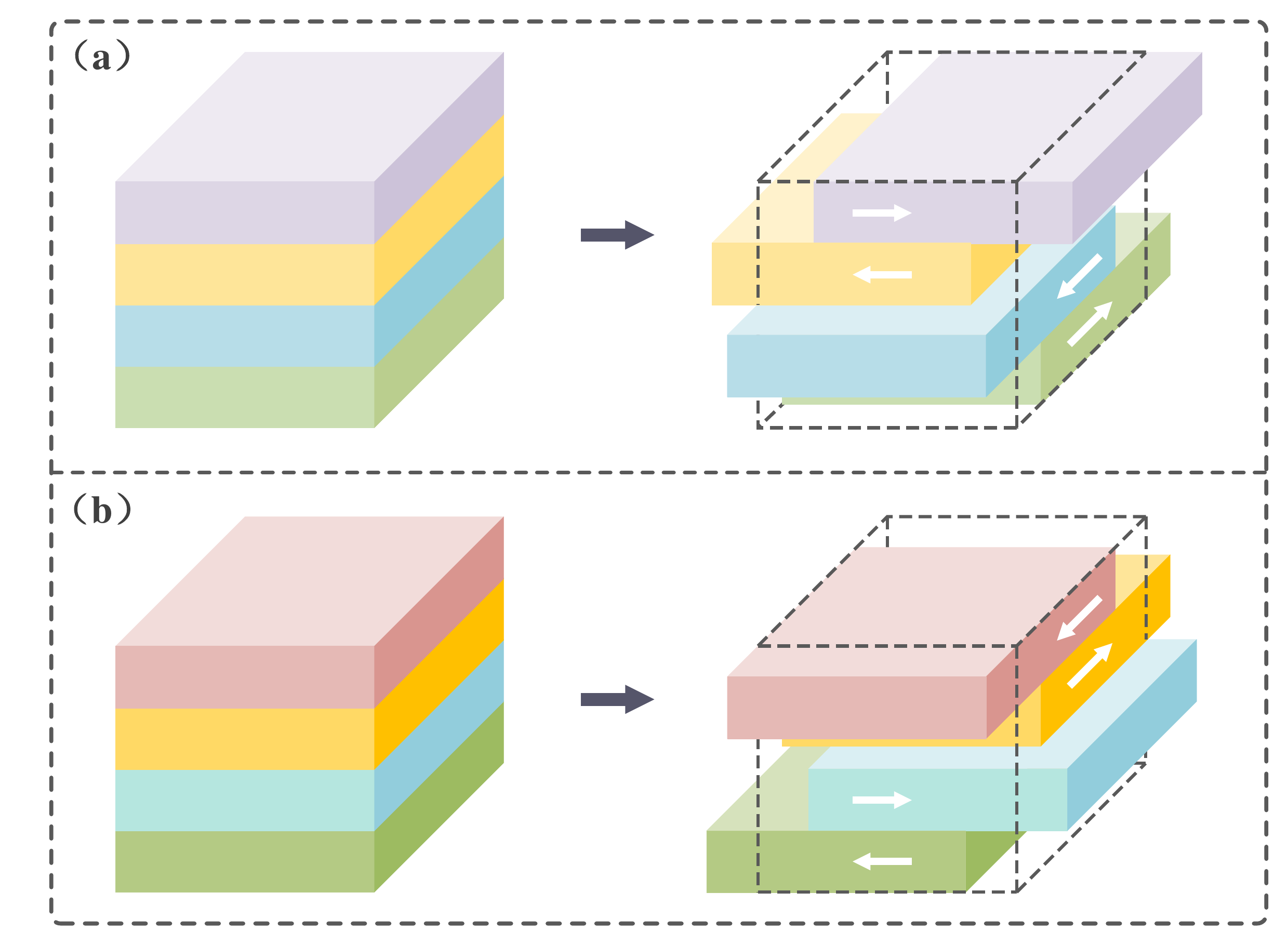}
    }
    \caption{The illustration of spatial shifts in the Spatial Shift Block. Panels (a) and (b) show two different shift methods applied to the height and width, differing in the order of the shifts.}
    \label{fig:shiftMLP}
\end{figure}

\noindent \textbf{Tokenized Shift Block.} The Tokenized Shift Block expands feature maps along the channel dimension, effectively capturing spatial and channel features, thereby improving the model's image segmentation performance. The structure of this module is shown in the top-left corner of \textbf{Figure} \ref{fig:LSU-Net}. This structure first processes the input using a 1×1 convolution layer to adjust the feature channels. Simultaneously, the initial input feature map is fed into the Spatial Shift Block \cite{S2MLPv2}, which further processes the input features through a tokenization mechanism. The Spatial Shift Block expands the feature map along the channel dimension and then splits the expanded feature map into three parts. While one part remains stationary, the other two undergo spatial shift operations in two directions, as shown in \textbf{Figure} \ref{fig:shiftMLP}. \textbf{Figure} \ref{fig:shiftMLP}(a) corresponds to Eq.\eqref{eq:shift_1}: 
\begin{flalign}
    \label{eq:shift_1}
    & X[2:h,:,1:c/4] \gets X[1:h-1,:,1:c/4]\nonumber \\
    & X[1:h-1,:,c/4+1:c/2] \gets X[2:h,:,c/4+1:c/2]\nonumber \\
    & X[:,2:w,c/2:3c/4] \gets X[:,1:w-1,c/2:3c/4]\nonumber \\
    & X[:,1:w-1,3c/4:c] \gets X[:,2:w,3c/4:c]
\end{flalign}

\noindent where, \(X\) represents a single sample, while \(h\), \(w\), and \(c\) represent height, width, and channels, respectively. And \textbf{Figure} \ref{fig:shiftMLP}(b) corresponds to Eq.\eqref{eq:shift_2}:
\begin{flalign}
    \label{eq:shift_2}
    & X[:,2:w,1:c/4] \gets X[:,1:w-1,1:c/4]\nonumber \\
    & X[:,1:w-1,c/4+1:c/2] \gets X[:,2:w,c/4+1:c/2]\nonumber \\
    & X[2:h,:,c/2:3c/4] \gets X[1:h-1,:,c/2:3c/4]\nonumber \\
    & X[1:h-1,:,3c/4:c] \gets X[2:h,:,3c/4:c]
\end{flalign}

Different spatial shift operations are performed on these segmented parts, which are then fused through a partitioned attention mechanism. The original feature map and the processed feature map are then fused to enhance feature representation, as shown in Eq.\eqref{eq:6}.


\begin{equation}
X_1 = X + Shift(X)
\label{eq:6}
\end{equation}

where, \(shift\) refers to the Spatial Shift Block. Next, a DWSConv layer is used, which reduces parameters and computational complexity while maintaining the effectiveness of convolution operations. This is followed by a GNorm layer to normalize the feature map, accelerating training and improving the model's generalization ability. 
\begin{equation}    
X_{2} = {Conv}_{1 \times 1}\left( GNorm\left( DWSConv\left( X_{1} \right) \right) \right)
    \label{eq:7}
\end{equation}

Subsequently, a 1×1 convolution layer with GELU activation function is applied to further perform nonlinear transformation and feature extraction. Finally, another 1×1 convolution layer processes and fuses the features, outputting the final feature map. The process is calculated as Eq.\eqref{eq:8}.
\begin{equation}    
Y = {Conv}_{1 \times 1}\left( {GELU\left( X \right.}_{2} \right)) + {Conv}_{1 \times 1}(X)
    \label{eq:8}
\end{equation}

The downsampling module of the Tokenized Shift Block adds an Overlap Patch Embedding layer at the input end. This layer divides the input image into small patches and embeds representation, preserving more local information and enhancing the richness of feature representation.

\noindent \textbf{Loss Function.} We propose Multi-scale Deep Loss (MDL), which balances class features and reduces fuzzy edge influence to enhance accuracy. MDL downsamples the mask at different scales, assigning each U-Net level a corresponding mask. The loss is calculated from segmentation results, with each level using a distinct loss function. Finally, an Automatic 

\noindent Weighted Loss (AWL) \cite{AWL} function combines the multi-level losses, updating weights according to Eq.\eqref{eq:9}.

\begin{equation}
    L_{s} = {\sum\limits_{i=1}^{n}{\frac{1}{2\sigma_{i}^{2}}}L_{i}} + {\sum\limits_{i=1}^{n}{ln\left( {1 + \sigma_{i}^{2}} \right)}}
    \label{eq:9}
\end{equation}
where, \(L_{s}\) is the loss, \(n\) is the total number of layers in the multiscale output, \(L_{i}\) denotes the loss at different scales, and \(\sigma_{i}\) represents the adaptive weight of the loss function at different layers. The term \(ln\left( {1 + \sigma_{i}^{2}} \right)\) restrains excessive weight parameter enlargement, preventing extreme values during optimization. This approach enhances loss function stability and allows the model to automatically adjust the weights of different losses based on training data.

\section{EXPERIMENTS}
\label{sec:experments}
In this section, we validate our proposed network architecture on the UWMGI \cite{uw-madison-gi-tract-image-segmentation} and MSD Colon \cite{MSD} datasets. First, we outline some implementation details. Then, we compare LSU-Net with other state-of-the-art architectures on the two datasets and analyze the experimental results.

\noindent \textbf{Implementation Details.} We develop LSU-Net using Pytorch \cite{pytorch} framework. All experiments are performed on a single NVIDIA GeForce RTX 3090 GPU. To evaluate our methods, we employ Mean Intersection over Union (mIoU), Dice similarity coefficient(DSC) as metrics. The mIoU and DSC are calculated as:

\begin{equation}
mIoU = \frac{1}{c}{\sum\limits_{i = 1}^{c}\frac{X \cap Y}{X \cup Y}} \label{eq:miou}  
\end{equation}

\begin{equation}
    DSC = \frac{1}{c}{\sum\limits_{i = 1}^{c}\frac{2\left| {X \cap Y} \right|}{|X| + |Y|}}
    \label{eq:dice}
\end{equation}

\begin{figure}[H]
    \centering
    \includegraphics[width=\linewidth]{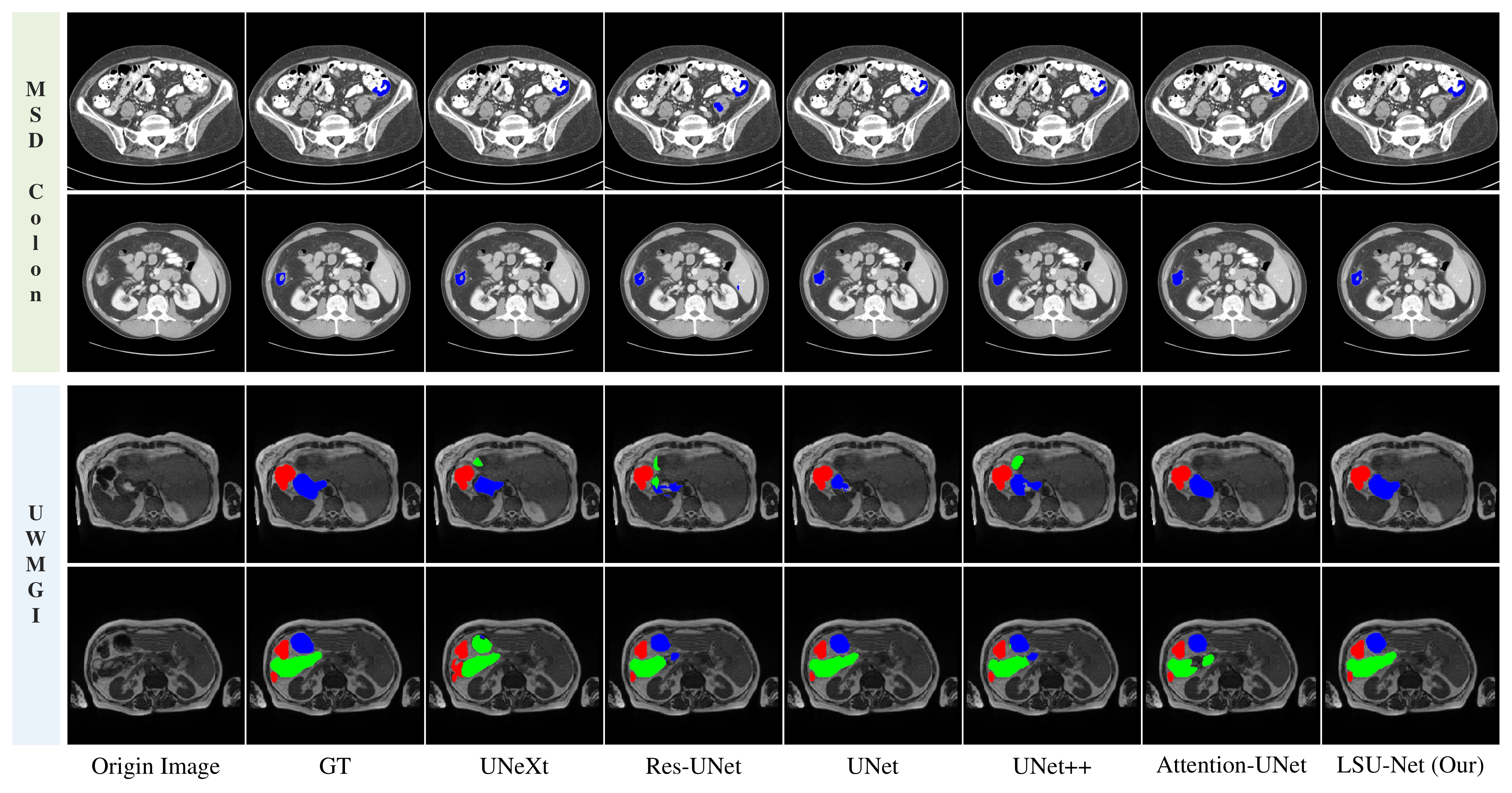}
    \vspace{-15pt}
    \caption{Visualization of segmentation results on the UWMGI and MSD Colon Cancer datasets. The above visualization shows MSD Colon Cancer results, while the bottom shows UWMGI results. From left to right: origin image, ground truth and the results of UNeXt\cite{UNext}, Res-UNet\cite{ResUNet}, UNet\cite{UNet}, UNet++\cite{UNet++}, Attention-UNet\cite{Attention-UNet}, and LSU-Net.}
    \label{fig:监督可视化}
\end{figure}

\begin{table}[t]
\centering
\caption{Comparative experimental results on the UWMGI dataset.}
\label{tab:supervised-UWMGI}
\vspace{0.2em} 
\resizebox{\columnwidth}{!}{
\begin{tabular}{l|c|c|c|c}
\toprule
Network & Param(M)$\downarrow$ & GFLOPs$\downarrow$ & mIoU(\%) & DSC(\%) \\
\midrule
Attention-UNet\cite{Attention-UNet} & 34.88 & 51.01 & 83.61 & 86.95 \\
Res-UNet\cite{ResUNet} & 13.04 & 52.13 & 79.06 & 82.70 \\
UNet++\cite{UNet++} & 9.16 & 26.50 & 83.33 & 86.65 \\
UNet\cite{UNet} & 7.85 & 10.73 & 84.02 & 87.41 \\
UNeXt\cite{UNext} & 1.47 & 0.39 & 82.29 & 86.02 \\
LSU-Net(Ours) & 1.08 & 1.10 & \textbf{86.04} & \textbf{89.43}\\
\bottomrule
\end{tabular}
}
\end{table}

\begin{table}[t]
\centering
\caption{Comparative experimental results on the MSD Colon Cancer dataset.}
\label{tab:supervised-MSD-Colon}
\vspace{0.2em} 
\resizebox{\columnwidth}{!}{
\begin{tabular}{l|c|c|c|c}
\toprule
Network & Param(M)$\downarrow$ & GFLOPs$\downarrow$ & mIoU(\%) & DSC(\%) \\
\midrule
Attention-UNet\cite{Attention-UNet} & 34.88 & 51.01 & 65.40 & 78.47 \\
Res-UNet\cite{ResUNet} & 13.04 & 52.13 & 53.88 & 69.56 \\
UNet++\cite{UNet++} & 9.16 & 26.50 & 67.11 & 80.05 \\
UNet\cite{UNet} & 7.85 & 10.73 & 69.63 & 81.92 \\
UNeXt\cite{UNext} & 1.47 & 0.39 & 72.39 & 83.84 \\
LSU-Net(Ours) & 1.08 & 1.10 & \textbf{72.79} & \textbf{84.05}\\
\bottomrule
\end{tabular}
}
\end{table}
\noindent where \(c\) denotes the number of classes, \(X\) and \(Y\) denote the predicted values and labels, respectively.

Images are normalized and resized to 224×224 for the UWMGI dataset and 512×512 for the MSD Colon Cancer dataset. Adam \cite{Adam} is used as the optimizer with an initial learning rate of 0.001, and CosineAnnealingLR \cite{SGDR} as the scheduler. The scheduler is configured with a maximum number of iterations and epochs, and a minimum learning rate of 1e-5. Both datasets are trained for 100 epochs, with batch sizes of 16 for UWMGI and 10 for MSD Colon Cancer. We evaluate our methods using Mean Intersection over Union (m-

\begin{table}[htbp]
\centering
\caption{Ablation studies examine the contributions of each module on the UWMGI dataset.}
\label{tab:ablation-supervised-uwmgi}
\vspace{0.2em} 
\resizebox{\columnwidth}{!}{
\begin{tabular}{c|c|c|c|c}
\toprule
Network & Params(M) & GFLOPs & mIoU(\%) & DSC(\%) \\
\midrule
w/o light conv block & 0.77 & 2.81 & 84.85 & 88.28 \\
w/o MDL & 1.08 & 1.1 & 84.81 & 88.25 \\
w/o tokenized shift block & 0.26 & 2.1 & 85.40 & 88.94 \\
Our Model & 1.08 & 1.1 & \textbf{86.04} & \textbf{89.43} \\
\bottomrule
\end{tabular}
}
\end{table}

\begin{table}[htbp]
\centering
\caption{Ablation studies examine the contributions of each module on the MSD Colon Cancer dataset.}
\label{tab:ablation-supervised-MSD}
\vspace{0.2em}
\resizebox{\columnwidth}{!}{
\begin{tabular}{c|c|c|c|c}
\toprule
Network & Params(M) & GFLOPs &  mIoU(\%) & DSC(\%)\\
\midrule
w/o light conv block & 0.77 & 2.81 & 71.91 & 83.42 \\
w/o MDL & 1.08 & 1.1 & 72.51 & 83.81 \\
w/o tokenized shift block & 0.26 & 2.1 & 70.81 & 82.73 \\
Our Model & 1.08 & 1.1 & \textbf{72.79} & \textbf{84.05} \\
\bottomrule
\end{tabular}
}
\end{table}

\noindent IoU) and Dice similarity coefficient (DSC).

\noindent \textbf{Experiment Results.} The results in Table \ref{tab:supervised-UWMGI} show that LSU-Net outperforms other models on the UWMGI dataset. LSU-Net surpasses the baseline UNeXt with a 3.75\% mIoU and 3.41\% DSC improvement. Compared to Res-UNet, LSU-Net not only achieves a 6.98\% mIoU and 6.73\% DSC gain but also reduces parameters and computation by 12.1x and 47.4x, respectively. LSU-Net also outperforms UNet++ with a 2.71\% mIoU and 2.78\% DSC improvement and exceeds Attention-UNet while reducing parameters and computation by 32.3x and 46.4x, respectively. On the MSD Colon Cancer dataset, as shown in Table \ref{tab:supervised-MSD-Colon}, LSU-Net balances performance and parameters, delivering excellent segmentation results. \textbf{Figure} \ref{fig:监督可视化} provides specific segmentation examples.

\noindent \textbf{Ablation Results.} We conducted ablation studies on the UWMGI and MSD Colon Cancer datasets to assess the impact of each module on supervised learning. Table \ref{tab:ablation-supervised-uwmgi} shows that incorporating MDL significantly improves performance, with a 1.23\% mIoU and 1.18\% DSC boost on UWMGI compared to the model without MDL. Removing other modules similarly reduces performance. Table \ref{tab:ablation-supervised-MSD} shows this effect is also observed in the MSD Colon Cancer dataset. The results indicate varying contributions of each module, with MDL having a more pronounced effect on the multi-class UWMGI dataset compared to the single-class MSD Colon Cancer, suggesting the need for dataset-specific fine-tuning. Ablation of the Light Conv Block and Tokenized Shift Block shows both modules enhance performance across different datasets.

\section{CONCLUSION}
\label{sec:conclusion}
In this paper, We propose LSU-Net, a lightweight abdominal organ segmentation network that uses feature space shift and multi-scale loss to enhance the recognition of organ regions, while employing lightweight convolutions to reduce model parameters. We conduct extensive experiments on multiple abdominal datasets, demonstrating that the proposed LSU-Net surpasses state-of-the-art segmentation models and advances cutting-edge performance. Our model holds significant potential for clinical applications, such as quantifying lesion areas and monitoring disease progression.





\bibliographystyle{IEEEbib}
\bibliography{Mybib}

\end{document}